\documentclass[fleqn,twoside]{article}
\usepackage{espcrc2,psfig}

\newlength{\figwidth}
\newcommand{\simgt}{\stackrel{>}{\scriptstyle\sim}}

\title{Anisotropic lattice with nonperturbative accuracy%
       \thanks{Poster presented by H. Matsufuru}}

\author{
Hideo Matsufuru\address{%
    High Energy Accelerator Research Organization (KEK),
    Tsukuba 305-0801, Japan \vspace{-0.25cm} },
Hidenori~Fukaya\address{%
    Yukawa Institute for Theoretical Physics, Kyoto University,
    Kyoto 606-8502, Japan \vspace{-0.25cm} },
Masanori Okawa\address{%
    Department of Physics, Hiroshima University,
    Higashi-hiroshima 739-8526, Japan \vspace{-0.25cm}},
Tetsuya~Onogi$^{\rm b}$,
and
Takashi~Umeda$^{\rm b}$ }

\begin{document}

\begin{abstract}
We determine the nonperturbative
anisotropic parameter of the gauge action in the quenched
approximation with less than 1\% 
accuracy using the Sommer scale measured by the L\"uscher-Weisz 
algorithm or smearing technique.
We also study the nonperturbative O(a)-improvement of the quark 
action.
The bare quark anisotropy is determined using the masses from
the temporal and spatial directions.
For the determination of the $O(a)$ improvement coefficients, 
we apply the Schr\"odinger functional method.
\end{abstract}

\maketitle

\section{Introduction}
  \label{sec:introduction}

Anisotropic lattices whose temporal lattice spacing $a_\tau$
is finer than the spatial one $a_\sigma$ have become a powerful
tool in various subjects of lattice QCD simulations.
Among these applications, computations of heavy-light
matrix elements \cite{Aniso01a,Aniso01b,Aniso02a,Aniso02_fD}
require the most accurate parameter tuning,
which should be performed nonperturbatively.
In this paper, we report the status of our project to develop
the anisotropic lattice framework for such precision computations
with accuracy of a few percent level \cite{Aniso04}.

Here we briefly summarize our strategy.
For precise computations of heavy-light matrix elements,
we need a framework of the heavy quark in which one should be
able to
(i) take the continuum limit,
(ii) compute the parameters in the action and the 
     operators nonperturbatively,
(iii) and compute the matrix elements with a modest computational
      cost.
The anisotropic lattice is a candidate of such framework, if
a method which fulfills the above condition (ii) is provided.
Our expectation is that on anisotropic lattices the mass dependence
of the parameters becomes so mild that one can adopt
coefficients determined nonperturbatively at massless limit.
We also need to control all the systematic errors in the
continuum extrapolations.
Feasibility studies performed so far for the level of $O(10\%)$
computations are encouraging for further development
\cite{Aniso02a,Aniso02_fD}.
We therefore investigate calibration procedures at
the accuracy less than one percent, both for the gauge and
quark actions in the quenched approximation
at $\xi=4$ \cite{Aniso04}.

\section{Calibration of gauge field}
\label{sec:calib_gauge}

To achieve a few percent accuracy in the final results,
the tuning of parameters must be performed at much less than
this accuracy.
As the goal of present work, we intend to determine the
anisotropy parameters at $O(0.2\%)$ level.
The elaborated work by Klassen \cite{Kla98},
the $O(1\%)$ level calibration for the Wilson action,
is therefore no longer meets the present condition.
For more precise calibration of the gauge field,
we need to measure the static quark potential very
accurately.
For this purpose, we adopt the L\"uscher-Weisz noise reduction
technique \cite{LW01} as well as the standard smearing technique
while applied in the anisotropic plane.
The former method can drastically reduce the statistical errors
while requires larger memory resources than the latter.

We define the renormalized anisotropy $\xi_G$ through
the hadronic radii $r_0$ measured in the coarse and fine
directions.
Since we carry out the continuum extrapolation in terms of the lattice
scale set by $r_0$, the renormalized anisotropy is kept
fixed during the extrapolation.
This avoids the systematic uncertainties
due to the anisotropy which may remain in the continuum limit.

\begin{figure}[tb]
\vspace*{-0.2cm}
\psfig{file=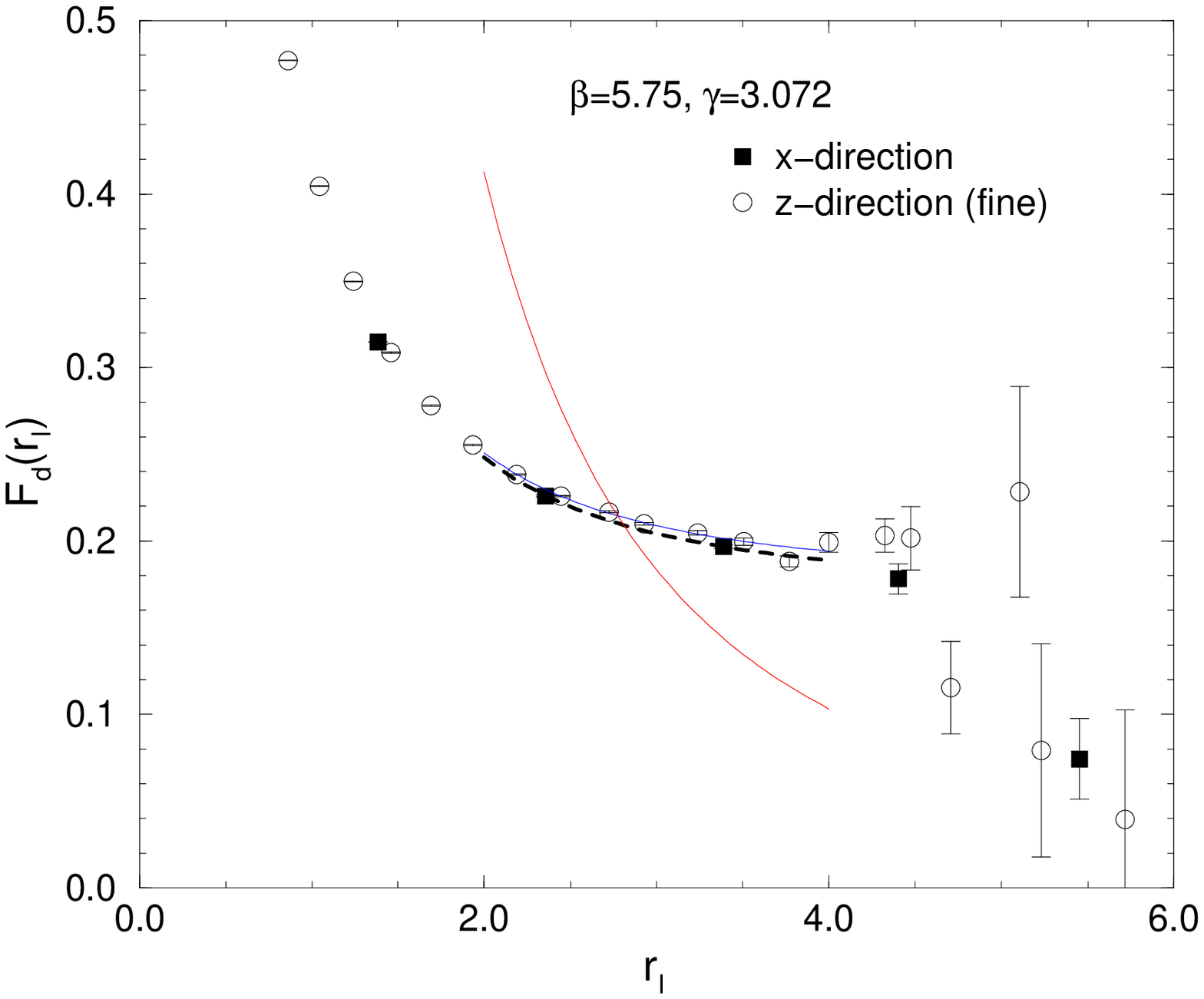,width=\figwidth}
\vspace{-0.8cm}
\psfig{file=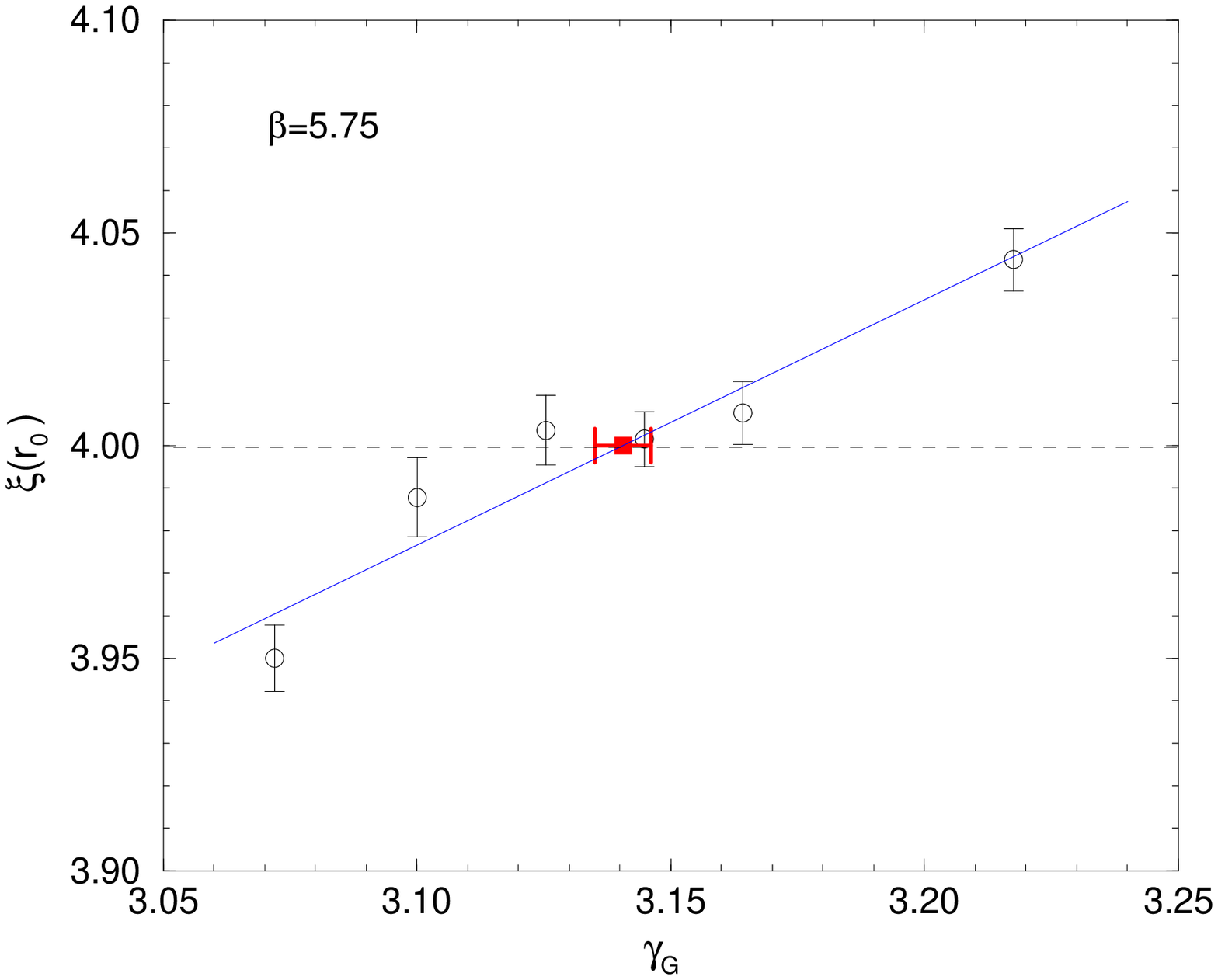,width=\figwidth}
\vspace{-1.4cm}
\caption{Determination of the gauge field anisotropy
at $\beta = 5.75$.}
\label{fig:gauge1}
\vspace{-0.8cm}
\end{figure}

Figure~\ref{fig:gauge1} shows a result of calibration
at $\beta=5.75$.
The top panel shows the result for the static potential
determined with the L\"uscher-Weisz technique at
$\gamma_G=3.072$.
The renormalized anisotropy $\xi_G$ is determined with
0.2\% accuracy.
A linear fit of the results at several values of $\gamma_G$
determines $\gamma_G^*$ for which $\xi_G=4$ holds
with 0.2\% accuracy as displayed in the bottom panel.

\begin{figure}[tb]
\vspace*{-0.2cm}
\psfig{file=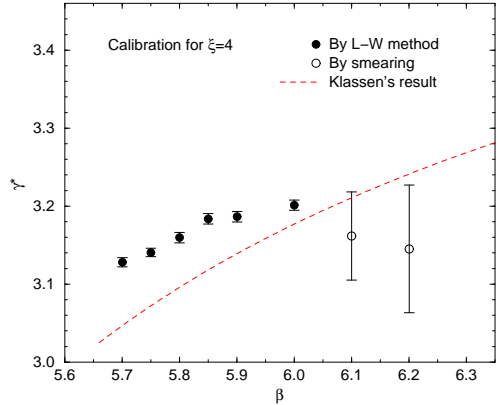,width=\figwidth}
\vspace{-1.4cm}
\caption{Result of calibration of gauge field.}
\label{fig:gauge2}
\vspace{-0.4cm}
\end{figure}

In Figure~\ref{fig:gauge2}, the results at several values of
$\beta$ are collected.
Although the precisions of the result with potential determined
with the standard smearing technique are still not enough,
sufficient precision is achieved at $\beta\leq 6$
where we used the L-W method.
Improvement of calculation at $\beta > 6$
and global fit analysis are in progress.

\section{Calibration of quark field}
\label{sec:quark_gauge}

Our heavy quark formulation basically follows the
Fermilab approach \cite{EKM97} but is formulated on the anisotropic
lattices \cite{Aniso01a,Ume01}.
The quark action is represented as
\begin{eqnarray}
 S_F &=& \sum_{x,y} \bar{\psi}(x) K(x,y) \psi(y),\\
 K(x,y) \!\!\!&=&\!\!\!
 \delta_{x,y}
   - \kappa_{\tau} \left[ \ \ (1-\gamma_4)U_4(x)\delta_{x+\hat{4},y} \right.
 \nonumber \\
 & &  \hspace{1cm}
      + \left. (1+\gamma_4)U_4^{\dag}(x-\hat{4})\delta_{x-\hat{4},y} \right]
 \nonumber \\
 & & \hspace{-0.2cm}
    -  \kappa_{\sigma} {\textstyle \sum_{i}}
         \left[ \ \ (r-\gamma_i) U_i(x) \delta_{x+\hat{i},y} \right.
 \nonumber \\
 & & \hspace{1cm}
     + \left. (r+\gamma_i)U_i^{\dag}(x-\hat{i})\delta_{x-\hat{i},y} \right]
 \nonumber \\
 & & \hspace{-0.2cm}
    -  \kappa_{\sigma} c_E
             {\textstyle \sum_{i}} \sigma_{4i}F_{4i}(x)\delta_{x,y}
 \nonumber \\
 & & \hspace{-0.2cm}
    - r \kappa_{\sigma} c_B
             {\textstyle \sum_{i>j}} \sigma_{ij}F_{ij}(x)\delta_{x,y},
 \label{eq:action}
\end{eqnarray}
where $\kappa_{\sigma}$ and  $\kappa_{\tau}$ 
are the spatial and temporal hopping parameters, $r$ the spatial
Wilson parameter and  $c_E$ and $c_B$  the clover coefficients.
For a given $\kappa_\sigma$, in principle, the 
four parameters $\gamma_F \equiv \kappa_{\tau}/\kappa_{\sigma}$,
$r$, $c_E$ and $c_B$ should be tuned.
We can set $r=1/\xi$ without loss of generality
\cite{Aniso01a,EKM97}.

\begin{figure}[tb]
\vspace*{-0.2cm}
\psfig{file=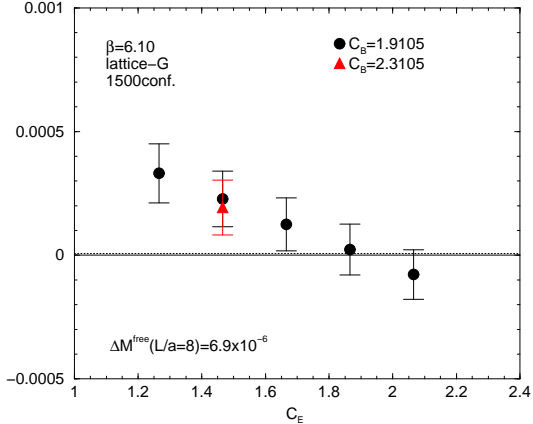,width=0.92\figwidth}
\vspace{-1.1cm}
\caption{Tuning of $c_E$ at $\beta=6.1$.}
\label{fig:quark1}
\vspace{-0.5cm}
\end{figure}

\begin{figure}[tb]
\vspace*{-0.2cm}
\psfig{file=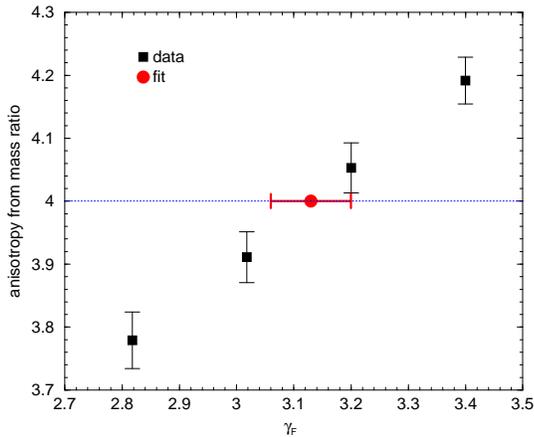,width=0.92\figwidth}
\vspace{-1.1cm}
\caption{Determination of $\gamma_F^*$ at $\beta=5.75$.}
\label{fig:quark2}
\vspace{-0.5cm}
\end{figure}

We must calibrate
$\gamma_F$, $c_E$, and $c_B$ to the level which enables
computations of matrix elements within a few percent accuracy.
We also need to perform the nonperturbative renormalization
of the operators such as the heavy-light axial current.
The nonperturbative renormalization technique \cite{NPR}
is one of the most powerful methods to perform such a program.
Following our strategy, this technique can also be applied with
a little modification for the anisotropic lattice.

We perform the calibration of $\gamma_F$, $c_E$, $c_B$,
and the renormalization coefficients of the axial current along
the following steps.
(1) Tuning of $c_E$ by Schr\"odinger functional method.
(2) Calibration of $\gamma_F$ (and $c_B$ if possible) by requiring
 the physical isotropy conditions for $m_{PS}$ and $m_V$
 in the coarse and fine directions on lattices
 with $T$,$L$ $\simgt$ 2 fm.
(3) Determination of $c_B$, $\kappa_c$ and the renormalization
coefficients of the axial current by Schr\"odinger functional method.
We also need to verify that the systematic errors are under control
by calculating the hadron spectra and the dispersion relations
and by taking the continuum limit.
It is also necessary to verify that the tuned parameters
in the massless limit is also available in the heavy quark mass
region.

To verify the feasibility of the step (1), we determine
$c_E$ with fixed values of $\gamma_F$ and $c_B$.
At $\beta=9.5$, the method is successfully applicable,
and the result for $c_E$ is close to the 1-loop mean-field value.
Figure \ref{fig:quark1} shows the result at $\beta=6.1$
($a_\sigma\simeq 2$ GeV).
The tuned value of $c_E$ is obtained as the value at which
$\Delta M$, the difference of the quark mass defined through
the axial Ward identity under different kinematical conditions,
vanishes up to $O(a^2)$ effects.
The result of $c_E$ is larger than the tadpole improved tree level
value.
It is also found that $\Delta M$ is not sensitive to the change of
$c_B$.

Figure \ref{fig:quark2} shows the result for the step (2)
at $\beta=5.75$ on a $12^2\times 24 \times 96$ lattice.
The values of $\xi_F$ is determined from the meson masses in
the fine and coarse directions.
We note that the result for $\gamma_F^*$ is consistent with that
from the dispersion relation \cite{Aniso01b}.
At this stage, the precision of $\gamma_F^*$ is still not
sufficient, while several techniques to reduce the statistical noise
are yet to be tested.
For the determination of $c_B$, the ratio of the hyperfine
splittings in the fine and coarse directions is not feasible
because of large statistical noise.
Other procedures, such as Schr\"odinger functional method with
boundaries in the coarse direction, are under investigation.

\end{document}